\documentclass[final,1p,times]{elsarticle}

\usepackage{hyperref}
\usepackage{geometry}

\journal{Applied Intelligence} 

\usepackage{balance}
\usepackage{listings}
\usepackage{color}

\definecolor{dkgreen}{rgb}{0,0.6,0}
\definecolor{gray}{rgb}{0.5,0.5,0.5}
\definecolor{mauve}{rgb}{0.58,0,0.82}

\makeatletter
\def\lst@makecaption{%
  \def\@captype{table}%
  \@makecaption
}
\makeatother

\graphicspath{{img/}}

\usepackage{url}
\usepackage{multirow}

\usepackage{amssymb}
\usepackage{amsmath}

\usepackage{array}
\usepackage{caption}
	\usepackage{graphicx}
\usepackage{subfigure}
	\usepackage{float}
	\usepackage{hyperref}
	\usepackage{amssymb}
	\usepackage{algorithm}
\usepackage{algorithmic}
	\usepackage{epsfig}
	\usepackage{epstopdf}
	\usepackage{lipsum}

\begin{document}

\begin{frontmatter}

\title{A Jumping Mining Attack and Solution}

\author{Muchuang Hu$ ^{1} $,
	Jiahui Chen$ ^{2}$,
	Wensheng Gan$ ^{3} $*,
and	Chien-Ming Chen$ ^{4} $}
\address{$ ^{1} $Department of Science and Technology, People's Bank of China Guangzhou, Guangzhou 510120, China} 
\address{$ ^{2} $School of Computer, Guangdong University of Technology, Guangzhou 510006, China}
\address{$ ^{3} $College of Cyber Security, Jinan University, Guangzhou 510632, China}

\address{$ ^{4} $College of Computer Science and Engineering, Shandong University of Science and Technology, Qingdao 266590, China}

\address{Email: moonchallenge@gmail.com, csjhchen@gmail.com, wsgan001@gmail.com, chienmingchen@ieee.org}

\cortext[cor1]{Corresponding author. Email: wsgan01@gmail.com} 

\begin{abstract}

Mining is the important part of the blockchain used the proof of work (PoW) on its consensus, looking for the matching block  through  testing a number of hash calculations. In order to attract more hash computing power, the miner who finds the proper  block can obtain some rewards. Actually, these hash calculations ensure that  the data of the blockchain is not easily tampered. Thus, the incentive mechanism for mining affects the security of the blockchain directly. This paper presents an approach to  attack against  the difficulty adjustment algorithm (abbreviated as DAA) used in blockchain mining, which has a direct impact on miners' earnings. In this method, the attack miner jumps between different blockchains to get more benefits than the honest miner who keep mining on only one blockchain. We build a probabilistic model to simulate the time to obtain  the next block at different hash computing power called hashrate. Based on this model, we analyze the DAAs of the major cryptocurrencies, including Bitcoin, Bitcoin Cash, Zcash, and Bitcoin Gold. We further verify the effectiveness of this attack called jumping mining through simulation experiments, and also get the characters for the attack in the public block data of Bitcoin Gold. Finally, we give an  improved DAA scheme against this attack. Extensive experiments are provided to support the efficiency of our designed scheme.

\end{abstract}


\end{frontmatter}


\section{Introduction}


In 2009, Nakamoto \cite{nakamoto2019bitcoin} firstly proposed the original concept of blockchain in the Bitcoin-based peer-to-peer electronic cash scheme. Since its release, blockchain has been extensively researched and developed globally, and its successful experience has attracted many organizational research on how to use blockchain technology in recent years. So far, there are more than 1,300 blockchain cryptocurrencies in the world, such as Bitcoin, Ethereum \cite{wood2014ethereum}, Ripple \cite{armknecht2015ripple}, etc. According to incomplete estimates, the cryptocurrency market is currently  worth more than 150 billion USD. There is no doubt that we should pay attention to the security of blockchain-based systems. How can these systems against the current or future attacks from the classical (non-quantum) and quantum adversaries is quite important.

The blockchain is essentially a distributed consensus storage system \cite{nakamoto2019bitcoin}, with consensus protocols between nodes to make agree on the contents of the storage. It can ensure that the ledger stored by each node in the distributed network are always consistent. Consensus protocols are, therefore, one of the key technologies in blockchain \cite{wang2020proof}. In fact, with the development of blockchain, many projects have proposed different consensus protocols, including proof of work (PoW) \cite{nakamoto2019bitcoin}, proof of stake (PoS) \cite{wood2014ethereum}, delegated proof of stake (DPoS) \cite{schuh2017bitshares}, practical Byzantine fault tolerance (PBFT)  \cite{castro2002practical}, etc. For the details of these consensus protocols, we recommend to refer the review paper \cite{mingxiao2017review}. The proof of work (PoW) used by Bitcoin is adopted by many public blockchain projects or systems. The principle of PoW is to achieve consensus by computing a mathematical problem. The miners who want to generate the next block in the blockchain to package new transactions must solve this problem.

Generally speaking, the computing problem is to calculate and find a proper hash value $\textit{Hash}(X)$ which is less than the target value \textit{PoW\_Target}, i.e. $\textit{Hash}(X)$ $\leq$ \textit{Pow\_Target}, where \textit{Hash} is a cryptographic hash function. Here $X$ is a random value determined by a nonce and the \textit{hash} value of the previous block. For example, there are fields including version, HashPreBlock \cite{nakamoto2019bitcoin}, HashMerkleRoot \cite{nakamoto2019bitcoin}, Time \cite{nakamoto2019bitcoin}, and Nonce \cite{nakamoto2019bitcoin} in the block head of Bitcoin. HashPreBlock is the hash value of the previous block. If the previous block value is modified, it can be easily verified by the hash value, which ensures that the historical block data is immutable. Bit stands for  the difficulty of mining. The calculated hash value must be less than or equal to this target hash value. Nonce is an random number. Mining is to modify the random number so that the hash value of the entire block can meet the target hash value. In fact, it is a contest with nodes involved in this puzzle competing against each other, and whoever finds the hash value smaller than \textit{PoW\_Target} firstly can generate the next block and get a reward. These nodes are the so-called miners. Algorithms that can be used as the hash functions $Hash$ include SHA256 \cite{courtois2014optimizing}, Scrypt \cite{watkins2017scrypt}, Ethash \cite{wiki2017ethash}, Cryptonight \cite{seigen2013cryptonight}, Equihash \cite{biryukov2017equihash}, X11 \cite{X11}, etc.

Generating a hash value is a random process, and the target value \textit{PoW\_Target} sets the difficulty for this computing problem. The task that looking for a $\textit{Hash}(X)$ $\leq$ \textit{Pow\_Target} is actually a probability problem. When the number of hash calculations that a miner can perform per unit time increases, the probability for him/her to find a matching hash grows. It means that the higher hash computing power (the so-called hashrate), the shorter time for getting the next block. Usually, a difficulty adjustment algorithm (DAA) \cite{nakamoto2019bitcoin} is used to ensure that the time for generating a block remains at a relatively stable value. When the average time for generating a block decreases, the difficulty can be increased by changing the \textit{PoW\_Target} value, and vice versa. The DAA addresses which time to adjust the \textit{PoW\_Target} value and how much to adjust. Thus, the DAA directly affects the time it takes for a miner to find the next block and get his/her mining rewards.

Hash calculation is actually the security mechanism to protect the blockchain. According to the longest chain consensus principle, if someone wants to tamper with the data in a blockchain, s/he needs to construct a blockchain that is longer than the existing one. To achieve this, s/he needs to recalculate the hash of the tampered block and all its previous blocks at a faster hashrate than the total value of the existing normal miners. If s/he has no more than 51\% of the hashrate of the entire network of the miners, it is theoretically impossible for her/him to do so. Therefore, it is important to attract more honest miners to participate in mining and to give the whole network a high level hashrate protection. As mentioned above, the DAA directly affects miners' benefit. If there are deficiencies with the DAA, it can cause fluctuations in the normal hashrate of the blockchain and threat the security of the blockchain.


The main goal of our study is to explore the interaction between PoW schemes and efficient DAA, so that future systems can achieve better fairness and better protection. We also intend to raise the awareness that a new attack are possible for PoW schemes, and that the assets protected by their deployments should be carefully valued. The methodology and contributions in this paper are as follows:

\begin{itemize}
	
	\item We firstly build a hashrate simulation model. The model can help us observe how the DAA adjusts the difficulty according the changes of the hashrate. At the same time, we analyze several DAAs of the major cryptocurrencies. We draw that these cryptocurrencies cannot react quickly or overreaction when the whole hashrate changes.
	
	\item By using a hashrate simulation model, we propose an attack method named jumping mining attack for different cryptocurrencies. The main idea of the jumping mining attack is to switch the hashrate between different cryptocurrencies used similar hash algorithm (e.g., SHA256, Scrypt, Ethash, Cryptonight, Equihash, X11, etc.), so that the attacker's benefit can be maximized. This attack leads to unstable hashrate, which directly affects the security of the blockchain.

	\item  What's more, in order to verify the effectiveness of the attack, we conduct a number of simulation attack experiments on three famous cryptocurrencies based on our attack method. The experimental results prove our scheme is effective, and we obtain further validation by the analysis of the public block data of Bitcoin Gold.
	
	\item Finally, we propose an improved DAA to resist the attack by summarizing the characteristics of the jumping mining attack. Similarly, we perform several experiments on the improved DAA and verify its effectiveness.
	
\end{itemize}

The rest of the paper is organized as follows: Section \ref{Related} describes the related work. In Section \ref{Model}, we describe the hashrate simulation model, analyze several DAAs of the major cryptocurrencies, and point out their weakness. In Section \ref{Attack}, we propose the jumping mining attack and validate it through some simulation experiments. In Section \ref{antiattack}, we provide an improved DAA against this attack. Finally, we conclude this paper in Section \ref{conclusion}.

\section{Related Works} \label{Related}


The blockchain technology has attracted much attention since it was first proposed in Nakamoto's original bitcoin paper  \cite{nakamoto2019bitcoin}. There are many use cases built around this technology. However, it also introduces a lot of speculation because the lack of legislations. Due to its openness and rapid economic growth, it has attracted many people to study on its security. Gervais et al. \cite{gervais2015tampering} showed that the scalability measures adopted by Bitcoin come at odds with the security of the system. Mayer et al. \cite{mayer2016ecdsa} discussed the security level of the signature scheme implemented in Bitcoin and Ethereum. After that, Moubarak et al. \cite{moubarak2018blockchain} also exposed  numerous possible attacks on the network. They evaluated blockchain security by summarizing its current state.

In these studies, some have focused on how miners mining strategies affect their income in the PoW blockchain. Nicolas et al. \cite{courtois2014subversive}  looked at the miner strategies with particular attention paid to subversive and dishonest strategies. After that, Kiayias et al. \cite{kiayias2016blockchain} studied the stochastic game that underlies these strategic. In the games, when the computational power of a miner is large, s/he deviates from the expected behavior, and other Nash equilibria arise.

DAA plays a key role in the mining process of PoW blockchains in order to maintain a consistent inter-arrival time between blocks. It is the core algorithm that influences the miner's strategies. Several studies have analyzed how DAA affects mining. Aggarwal et al. \cite{aggarwal2019structural} compare the equilibrium behavior of miners between Bitcoin's DAA and Bitcoin Cash's \cite{BitcoinCash}  emergency difficulty adjustment algorithm. Following Bitcoin Cash, considerable effort has been devoted to improve the DAA of  PoW blockchain. Kraft \cite{kraft2016difficulty} and Fullmer \cite{fullmer2018analysis} proposed an alternative DAA.

In addition to affecting the miner's income, DAA is more closely related to the security of the blockchain. Since Bitcoin was split into  Bitcoin and Bitcoin Cash (BCH) in August 2017, the miners had a choice between different  cryptocurrencies because they have compatible proof-of-work algorithms. There are several attacks focus on the famous cryptocurrencies. A double-spend attack through the hashrate leasing market was proposed by  Budish \cite{budish2018economic} in 2018. Biryukov et al. \cite{biryukov2019privacy} analyzed two privacy and security issues for the privacy-oriented cryptocurrency Zcash. They introduced two  attacks called Danaan-gift attack and Dust attack. After that, Auer \cite{auer2019beyond}  showed that in the long run PoW's security will mainly come from transaction fees.  In future research, we can test whether the theoretical analysis of other attacks can mitigate the observation of the impact on costs.

\section{Our Hashrate Simulation Model}\label{Model}

In this section, we give the simulation model to observe the reflection of the changes of the network hashrate. There are at least two limitations to analyze how the DAA of the public blockchain project works when the hashrate change. On the one hand, it takes a lot of time to  generate enough block data for analysis even though we test it on test-net. On the other hand, it is hard to get enough hashrate for test. Therefore, an effective model can greatly improve the efficiency of the analysis. The simulation code is available in Github\footnote{\url{https://github.com/humuchuang/jumping-mining}}.

\subsection{Hashrate Simulation Algorithm}

In general, mining is like a puzzle game. Firstly, we let the input be $X$ and the output be \textit{Hash}($X$) where $Hash$ is cryptographic hash function. Then the puzzle is to find an answer $X$ where  the value of \textit{Hash}($X$) is less than a specified target value \textit{PoW\_Target}. Take Bitcoin as an example, the hash function is SHA256. The miners apply a 256-bit cryptographic hash algorithm \cite{hashcash2019} to an 80-byte block header and an Nonce. The puzzle is solved if the resulting value is less than a known target value \textit{PoW\_Target}  where  0 $<$ \textit{PoW\_Target} $<$ \(2^{256}\) and \textit{Hash}($X$) $\leq$ \textit{PoW\_Target}. Here the input $X$ contain the 80-byte block header and the Nonce. To find a suitable $X$ is a process of exhaustive test by raising the value of the Nonce. Due to the randomness of the output of the hash function, finding a suitable input value $X$ is actually a probabilistic process. The more the hash calculation is performed, the greater the probability of finding a suitable solution. The hashrate of a miner actually refers to how many times he can perform hash calculations per unit time. The higher hashrate a miner has, the easier it is to find a suitable block. Thus, we can build a probabilistic model to simulate the time for generating a block at different hashrate.

The probabilistic derivation process of our hashrate simulation model is presented below. Similarly, we still take Bitcoin as an example. As we know, the specified target value \textit{PoW\_Target} of Bitcoin is a 256-bit number. The maximum value of \textit{PoW\_Target} denoted as \textit{Max\_Target} is \(2^{256}\). Usually, the blockchain will have an initial target value to limit the minimum mining difficulty.Taking Bitcoin as an example again, the limit target value is \(2^{224}\). Each different cryptocurrency sets different limit target values based on different  network hashrate. Here we assume that \textit{PoW\_Target}  equals  \(2^{248}\). For ease of understanding how to calculate the probability of finding a suitable answer, the assumed \textit{PoW\_Target} can be represented in hexadecimal as  $\rm{0x00ffffffffffffffffffffffffffffffffffffffffffffffffffffffffffffff}$.
The probability to get the suitable $X$ in a hash calculation is $P$ = $ \frac{2^{248}}{2^{256}}$. Let $D$ be the average number of hash calculations required to find $X$. Because of finding  a hash value smaller than the target value \textit{PoW\_Target} is an independent repeated probability event, we have  $D$ = $\frac{1}{P}$ = $\frac{2^{256}}{2^{248}}$ = 256.

Generally, the difficulty to find a matching hash value can be defined as follow:
\begin{equation}
D = \frac{Max\_Target}{PoW\_Target},
\end{equation}
where \textit{Max\_Target} indicates the maximum value of \textit{PoW\_Target} when solving the hash problem. Let $m$ be the number of leading zero of \textit{Max\_Target}, i.e., assume that the maximum target value for the main network of Bitcoin is $\rm{0x00000000ffffffffffffffffffffffffffffffff}$, $m$ = 8. Then the difficulty can be defined as follow:
\begin{equation}
D = \frac{2^{256-m}}{PoW\_Target }.
\end{equation}

We assume the leading zero $m$ = 0. The meaning of this difficulty $D$ is the number of hashes to be calculated. For example, if $D$ = \(2^{40}\), then it is necessary to find a hash value with 40 leading zeros. The probability of  matching the target in one hash calculation is $P$. Hence, we have
\begin{equation}
P = \frac{1}{D}.
\end{equation}
Then the probability that the $n$-th time the target value happens to be found is \( p*(1-p)^{n-1}\), namely the first $n-1$ times all fail and the last time succeeds. Given the difficulty, the probability cumulative function $P(n)$ represents the probability of finding a target block hash for the first $n$ times, as shown below:
\begin{equation}
P(n) = \sum\limits_{k = 1}^n { p*(1-p)^{n-1}} = p* \frac{{1-(1-p)^n}}{1-(1-p)} = 1-(1-p)^n.
\end{equation}

According the above derivation, the simulation for the $n$-th time the target value happens to be found can be conducted as the follow steps.
\begin{itemize}
	\item {\bf I}. Firstly find a random number \textit{Rand} in a $0-1$ uniform probability distribution.
	
	\item {\bf II}. Set an inequality equation $P(n-1) < Rand < P(n)$.
	
	\item {\bf III}. Solve the value of $n$.
\end{itemize}

Hence, we have:
\begin{equation}
n = ceil(\frac{{1og(1-Rand)}}{1og(1-p)}).
\end{equation}
Assume the average time for the honest miner to generate a block is $T$ seconds, then the hashrate of the honest miner can be defined as $HR$ = $\frac{D}{T}$. The hashrate of a miner actually refers to how many times s/he can perform hash calculations per unit time. After getting the number of hash calculations $n$ which is required for matching the target, the time of this process can be calculated as:
\begin{equation}
solvetime = \frac{n}{HR}.
\end{equation}

Finally, we use the key hashrate simulation algorithm to obtain the solve time for generating a block at any given difficulty, as described in Algorithm \ref{OurSim}.

\renewcommand{\algorithmicrequire}{\textbf{Input:}}
\renewcommand{\algorithmicensure}{\textbf{Output:}}
\begin{algorithm}
	\caption{\textsf{GetSolveTime}(\(HR,Rand,D\))}
	\label{OurSim}
	
	\begin{algorithmic}[1]
		\REQUIRE \textit{HR}: the total hashrate of the blockchain;\\ \textit{Rand}: a random number generated in a 0-1 uniform probability distribution;\\ $D$: the current difficulty to generate a block.
		
		\ENSURE $ST$: the time it takes to generate the next block.
		
		\STATE Set the base line of the difficulty \(Lz=2^{40}\);  // Similar to the limit target value explained before, the baseline here also ensures  the minimum difficulty.
		\STATE Calculate the probability of success to match the target in one hash calculation $p$ = $\frac{1}{D*Lz}$;
		\STATE $ST$ = $n/HR$ = $ceil(\frac{{1og(1-\textit{Rand})}}{1og(1-p)}$)*$\frac{1}{HR}$.\\
		\STATE \textbf{return} \(ST\)
	\end{algorithmic}
\end{algorithm}


\subsection{Analysis on DAAs of Different Cryptocurrencies}\label{DAAs}

In this section, we mainly analyze the problems with the DAAs of several major cryptocurrencies that currently adopt the PoW mechanism.

As we can see from the Bitcoin core code \cite{BitcoinCore2019}, Bitcoin's DAA is not complicated. Bitcoin adjusts its difficulty every 2016 blocks. If the whole hashrate is stable,the average time for generating a block is 10 minutes. It means the difficulty adjusts  once per two weeks. When the adjustment cycle comes, the blockchain calculate the time for generating  the previous 2016 blocks. If the time is less than two weeks, then the difficulty of next block will be increased in proportion. Conversely, if it is greater than two weeks, the value needs to be reduced. For example, if the previous 2016 blocks was generated about one week, then the difficulty should be double in the next  adjustment cycle. In addition, the proportional limit for difficulty adjustment is $[0.25, 4]$ in order to avoid over regulation. Under normal circumstances, this strategy is relatively stable. However, when there are large fluctuations in the hashrate of the blockchain network, the response of this strategy is relatively lagged. For example, the attacker chooses to enter on a lower difficulty cycle, and since he has a relatively high hashrate, he can generate a lot of blocks quickly. When the adjustment cycle comes, the difficulty increases drastically and the attacker chooses to  exit. As a result, the honest miners who keep mining will mine at a higher difficulty for a long time causing severe delays in the generation of blocks.

Bitcoin Cash \cite{BitcoinCash} was started out as a hard fork project of Bitcoin with new features at the beginning. Bitcoin Cash improved its difficulty adjustment algorithm after the fork. Bitcoin Cash's DAA works on a similar principle to BTC, using $N$ previous blocks as a reference and adjusting in order to generate a block every 10 minutes. The difference is that the difficulty adjustment of Bitcoin Cash  is block-by-block, while the number of the past blocks $N$ for reference is 144 rather than 2016. Likewise, to prevent over-adjustment, Bitcoin Cash has a proportional limit for difficulty adjustment within $[0.5, 2]$. Although this DAA is more responsive to the variety in the hashrate of the blockchain, it also creates instability. Another new proposal points out that $N$ should be smaller in the strategy that uses $N$ past blocks as a reference for difficulty adjustment in order to more accurately reflect changes of the hashrate of the blockchain network over the recent period.

Zcash \cite{Zcash} and Bitcoin GOLD \cite{BTG} is the cryptocurrency which uses this proposal. The DAA of Zcash called Digishiled, where $N$ = 17. Under this scheme, the most recently generated block  reflect the current state of the  network's hashrate best. However, another better scheme is to set weights on the reference blocks. In Bitcoin Gold's latest difficulty adjustment algorithm \cite{BTGDAA2020}, the difficulty adjustment with weights is used. The newer is the block, the higher weight is set. Let $N$ be the number of the reference block and the current height of the blockchain be $h$. When we calculate the difficulty of the $h+1$ block, the $h$-th block is weighted $N$ while the $(h$-1)-th block is weighted $N-1$, and so on. The sum of the weight is  \(\sum\limits_{i = 1}^N{i} = \frac{N*(N+1)}{2}\). The average target for the past $N$ blocks is  \textit{avgTarget} = $\sum\limits_{i = 1}^{N}{\frac{\textit{Target}(h-i)}{h-i}}$. And the average generation time for $N$ blocks in the past is \textit{avgT} = $\sum\limits_{i = 1}^N{\frac{ST(h-i)}{i}}$. Suppose the expected time to generate a block is $T$. Hence, the target value of next block is  \textit{newTarget} = $ \frac{{\textit{avgTarget}}*{\textit{avgT}}}{T*{\textit{adjust}}}$, where  \textit{adjust} is an adjustment factor less than 1.

Although this strategy already seems relatively reasonable, it still has several problems. It does not adequately take into account the fact that hashrate may jumps in and out frequently causing fluctuates dramatically. When a large hashrate jumps out of the network, the difficulty of generating blocks will increase significantly. What's worse, because the network hashrate is not enough, the time delay to generate blocks is longer, making the difficulty of the next few blocks will be significantly reduced. The block difficulty drops too quickly will raise new problem. In the following section we can see that as long as the attacker has a certain scale of hashrate and chooses the right timing for her/his attack, s/he can still gain better than the honest miner under this difficulty adjustment algorithm.

\section{Jumping Mining Attack}\label{Attack}

In this section, we introduce a scheme of jumping mining between different cryptocurrencies that using the same hash function in its PoW consensus algorithm. Therefore, the miners using this method can obtain higher mining revenue than the miners who keep mining on one cryptocurrency. The  jumping mining actually damages the earnings of honest miners who keep mining on one cryptocurrency. In the future, it may cause miners unwilling to act as honest miners, resulting in fluctuations in the cryptocurrencies' computing hashrate. Hashrate plays an important role in ensuring the security of the blockchain. If the total hashrate of a certain cryptocurrency drops to a certain level, it may be subject to other attacks, such as 51\% attack. Therefore, this method of jumping mining against the problems of the difficulty adjustment algorithm is actually a serious attack. Its worst case may shake the foundation of the security of the blockchain and eventually lead to the demise of the cryptocurrency.

Further more, based on the hashrate simulation model in Section \ref{Model}, we ran a simulation experiment of the jumping mining attack. The experiments not only recorded the hashrate curve of the entire tested cryptocurrency during the test, but also logged the time when the attacker moved his hashrate into mining and out. Hence, we can intuitively observe the block difficulty value and duration of each attack by the attacker. Finally, we compared the average block generation time and mining benefits of honest miners with the attacker to determine whether the attack is effective.

\subsection{The Attack Method}

According to the difficulty of adjustment algorithm, the value of the block difficulty vary as the computing hashrate of the miners of the entire network. In the DAA as mentioned before, the speed to update the block difficulty  is either too slow or overreacted. For example, Bitcoin adjusts the difficulty every 2016 blocks. If there is a big change in the hashrate of the whole network, it cannot be adjusted in time, while the other DAA are overreacted. Our attack strategy is naive and we assume that the attacker has a certain computing hashrate denoted as \textit{HRAttacker}, and the hashrate of other honest miners on the cryptocurrency network is \textit{HRworker}. The attacker can randomly choose the time to join the mining, which is actually the attack time stamp. By observing the block difficulty sequence of the cryptocurrency network, the attacker start to  mine when the difficulty is at a lower level, and jump out when the block difficulty is adjusted to a certain higher level. In this way, the mining efficiency of the attacker is thus higher than the honest miners. The key steps of the jumping mining attack are described in Algorithm \ref{attackprocedue}. It is worth mentioning that the parameters settings of ``the difficulty threshold" can change according the block difficulty fluctuation curve to obtain the best attack results.


\begin{algorithm}
	\caption{\textsf{JMAttack}}
	\label{attackprocedue}
	\begin{algorithmic}[1]
		\STATE Set the base line of the difficulty \(\textit{BaseD}*Lz = 4*2^{40}\);
		\STATE Set the hashrate of the honest miners \textit{HRworker} = $\frac{\textit{BaseD}*Lz}{T}$, where $T$ is the expected average time for the honest miner to generate a block;
		\STATE Set \textit{HRAttackerMulti} = 1; // Note that \textit{HRAttackerMulti} can be changed as needed.
		\STATE Set the hashrate of the attack miners \textit{HRAttacker} = \textit{HRAttackerMulti} * \textit{HRworker};
		\STATE Set the difficulty threshold  \textit{AttackIn} = 0.95; // When  the difficulty of the block is 5 percent lower than the benchmark level,the attacker start to attack.
		\STATE Set the difficulty threshold for exiting an attack \textit{AttackOut} = 1.45; // When  the difficulty of the block is 45 percent higher than the benchmark level,the attacker quit the attack.
		\STATE Let \textit{Dseri} be the block difficulty sequence;
		\STATE Let \textit{STseri} be the time sequence of block generation;
		\STATE Set an attack flag as \textit{Attackposition} == 0;
		\FOR {$i$ = 1 to $n$, where $i$ is the block height}
		\IF {$\textit{Dseri}(i-1) <$ \textit{AttackIn} * \textit{BaseD} and \textit{Attackposition} == 0}
		\STATE \textit{Attackposition} = 1;
		\STATE  \textit{HRnow} = \textit{HRAttacker} + \textit{HRworker}; \textit{HRnow} is total hashrate of the entire blockchain network;
		\ELSIF { $\textit{Dseri}(i-1) > $ \textit{AttackOut} * \textit{BaseD} and \textit{Attackposition} == 1}
		\STATE \textit{Attackposition} = 0;
		\STATE  \textit{HRnow} = \textit{HRworker};
		\ENDIF
		\STATE  $\textit{Dseri}(i)$ = \textit{GetNextDifficulty}($\textit{Dseri}(i$-$N$:$i$-1), $\textit{STseri}(i$-$N$:$i$-1), $T$, $N$); // \textit{GetNextDifficulty} is the difficulty adjustment algorithm defined by different cryptocurrencies.
		\STATE  $\textit{STseri}(i)$ = $ \textit{GetSolveTime}(\textit{HRnow}, \textit{RndSeri}(i), \textit{Dseri}(i))$ \ref{OurSim};
		\ENDFOR
	\end{algorithmic}
\end{algorithm}


In Algorithm \ref{attackprocedue}, we choose the attack timing according to the difficulty threshold of the block.  Note that we can also use other conditions as the trigger for starting and exiting the attack. Considering that Bitcoin's adjustment strategy is different, we chose 2016 blocks as an attack cycle rather than the difficulty threshold. Regardless of the choice, our fundamental purpose is to mine at a lower level of block difficulty and to exit at a higher level. Thus, the jumping mining attack on Bitcoin should be the following details, as described in Algorithm \ref{attackprocedue2}.

\begin{algorithm}
	\caption{ \textsf{AttackOnBitcoin} }
	\label{attackprocedue2}
	\begin{algorithmic}[1]
		\STATE Set the base line of the difficulty \(\textit{BaseD}*Lz = 4*2^{40}\);
		\STATE Set the hashrate of the honest miners \textit{HRworker} = ${BaseD*Lz}{T}$, where $T$ is the expected average time for the honest miner to generate a block;
		\STATE Let \textit{Dseri} be the block difficulty sequence;
		\STATE Let \textit{STseri} be time sequence of block generation;
		\STATE Set an attack flag as \textit{Attackposition} == 0;
		\FOR {$i=1$ to $n$ , where $i$ is the block height}
		\IF {$(mod(i, 2016)$ == 0) and \textit{Attackposition} == 0}
		\STATE  \textit{Attackposition} = 1; // \textit{Attackposition} is the attack flag
		\STATE  \textit{HRnow} = \textit{HRAttacker} + \textit{HRworker}; // \textit{HRnow} is total hashrate of the entire blockchain network
		\ELSIF { $(mod(i,2016) == 0)$ and \textit{Attackposition} == 1}
		\STATE  \textit{Attackposition} = 0;
		\STATE  \textit{HRnow} = \textit{HRworker};
		\ENDIF
		\STATE  $\textit{Dseri}(i)$ = \textit{GetNextDifficulty}($\textit{Dseri}(i$-$N$:$i$-1), $\textit{STseri}(i$-$N$:$i$-1), $T ,N)$; // \textit{GetNextDifficulty} is the DAA of Bitcoin.
		\STATE  \textit{STseri}($i$) = \textit{GetSolveTimeFunc}(\textit{HRnow}, \textit{RndSeri}$(i), \textit{Dseri}(i))$ \ref{OurSim}.
		\ENDFOR
	\end{algorithmic}
\end{algorithm}


\subsection{Attack Results}
We conduct attack experiments on the DAA of each cryptocurrency introduced earlier. We separately simulate the situation where the attacker's computing hashrate is equal to three times that of an honest miner. In terms of the timing of the attack, in addition to Bitcoin, we choose to attack when the block difficulty is as 95 percent of the base difficulty, and to exit when the block difficulty reaches 1.45 times of the base difficulty. We attack Bitcoin according its adjustment cycle. Note that the parameters for choosing the timing can be selected based on the actual attack data. Our recommended strategy is to select a difficulty threshold that maximizes the time for low difficulty mining.

\begin{figure*}[ht]
	\centering
	\subfigure[Equal Hashrate]{
		\includegraphics[width=8cm]{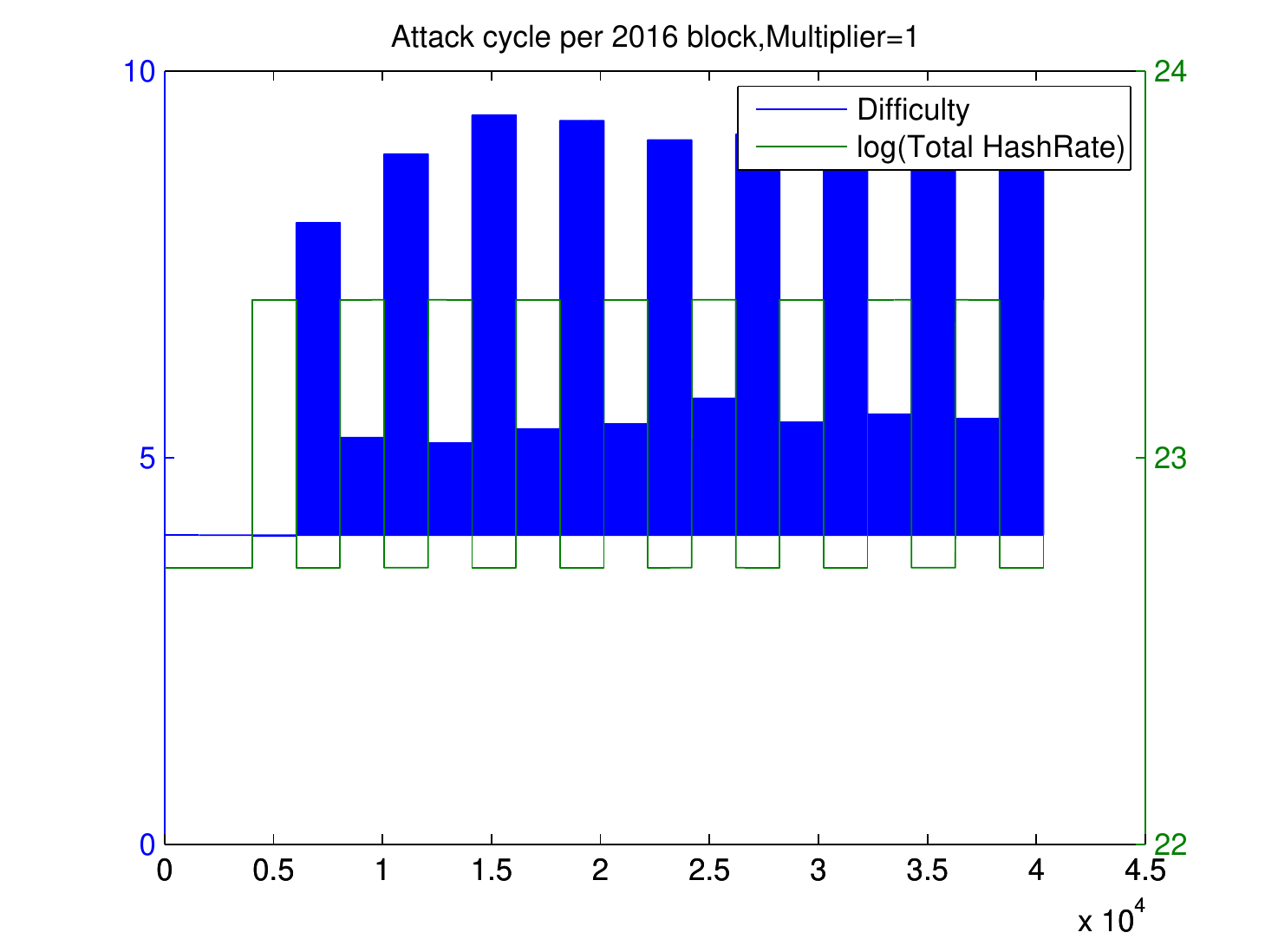}}
	\hspace{0in}
	\subfigure[Three Times Hashrate]{
		\includegraphics[width=8cm]{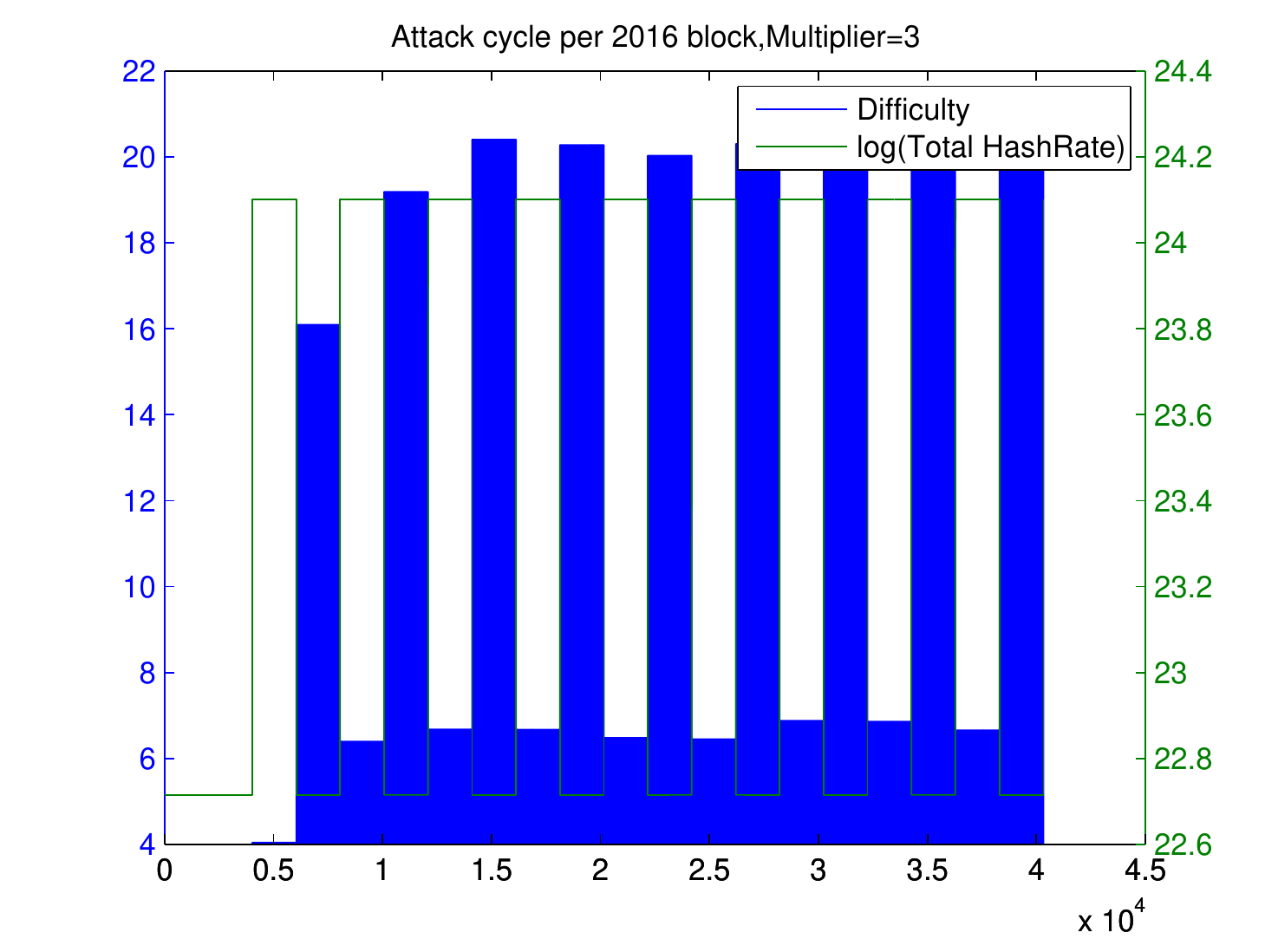}}
	\caption{\scriptsize{Simulation Results on Bitcoin}}
	\label{btc}
\end{figure*}

\textbf{Results on Bitcoin}. As shown in Figure \ref{btc}, the blue curve stands for the difficulty sequence, and the green one refers to the entire hashrate. The peak area of the orange curve is the attack period.

\begin{itemize}
	\item {\bf I}. When the attacker's hashrate is equal to the honest miner, the average time for the attacker to mine a block is 703.8s, while the honest miner's time requires 874.9s. The attacker's mining efficiency is 0.001421 while the honest one is 0.001143. Here we assume the  efficiency is equal to the average time to generate a block divided by the miner's hashrate. The benefits of the attacker  are significantly higher than the honest one.
	
	\item {\bf II}. When the attacker's hashrate is three times than that of the honest miner, the average time for the attacker to mine a block is 261.6s, while the honest miner's time takes 1472.4s. The attacker's mining efficiency is 0.001421, and the honest miner is 0.000679. With the increase of the attacker's computing hashrate, the attack efficiency is much higher.
	
\end{itemize}

\begin{figure*}[ht]
	\centering
	\subfigure[Equal Hashrate]{
		\includegraphics[width=8cm]{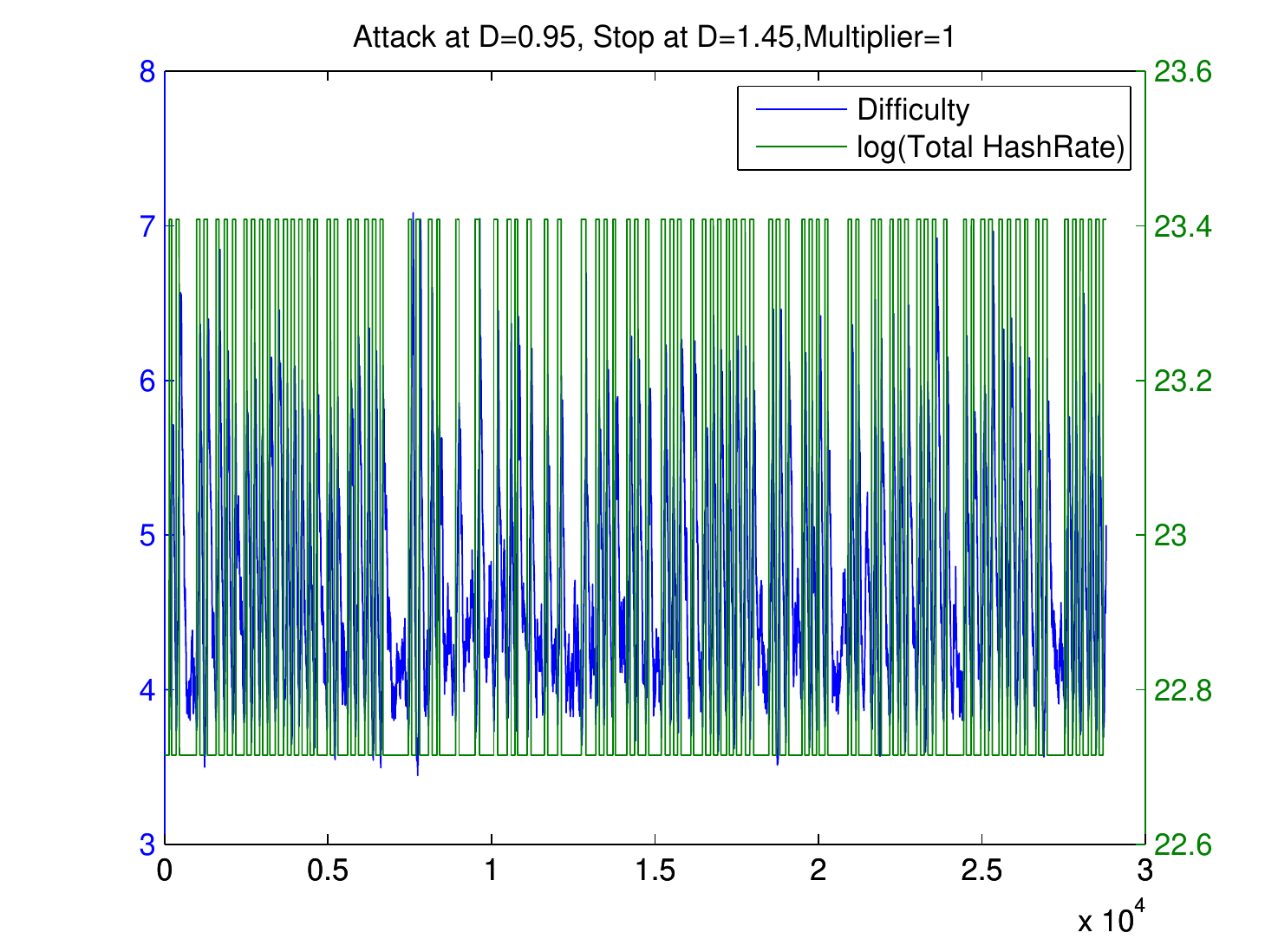}}
	\hspace{0in}
	\subfigure[Three Times Hashrate]{
		\includegraphics[width=8cm]{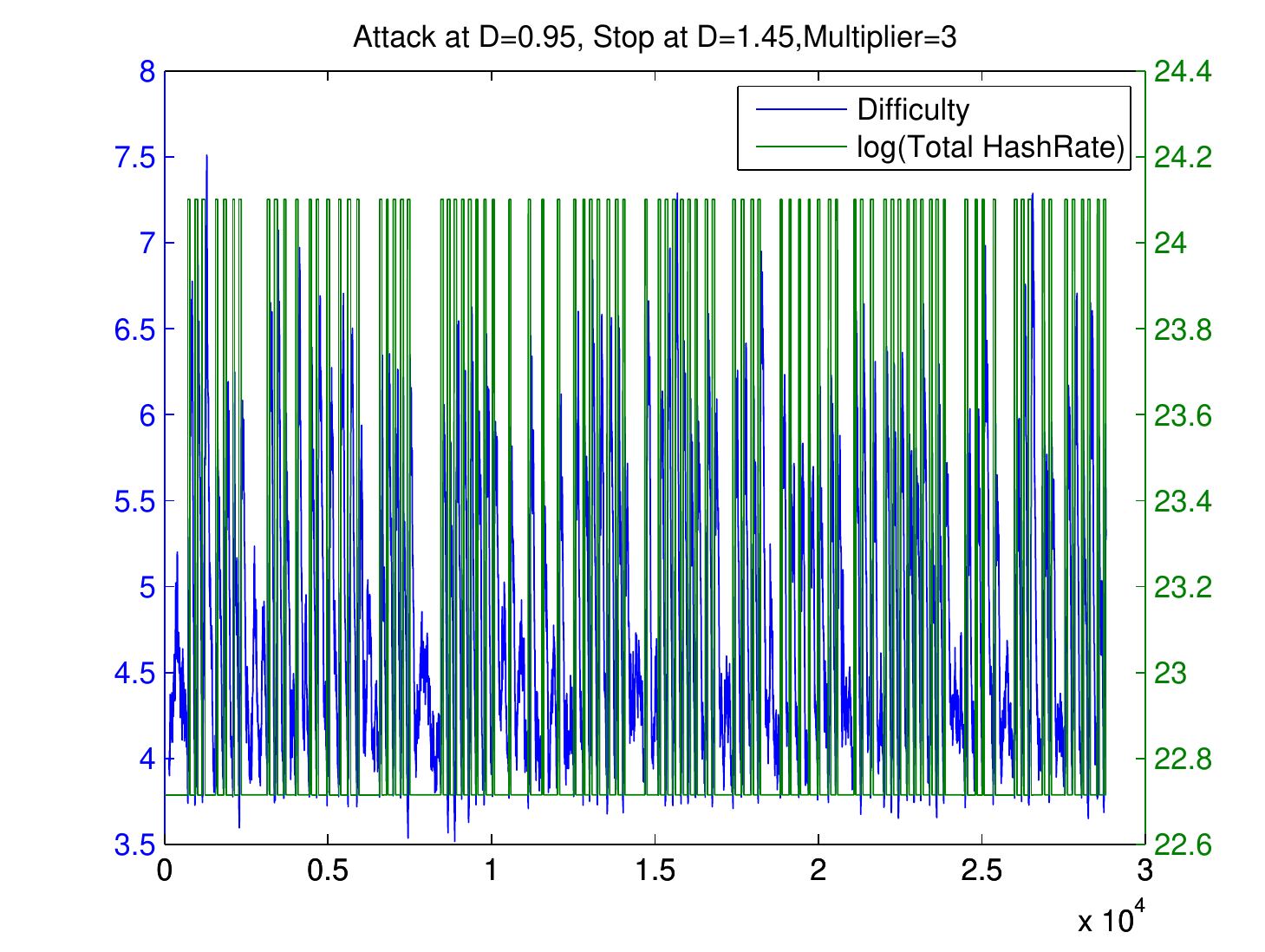}}
	\caption{\scriptsize{Simulation Results on Bitcoin Cash}}
	\label{bch}
\end{figure*}

\textbf{Results on Bitcoin Cash}. It can be seen from Figure \ref{bch} that the hashrate is more unstable and the fluctuation is more severe.

\begin{itemize}
	\item {\bf I}. When the attacker's hashrate is equal to the honest miner, the average time for the attacker to mine a block is 685.8s, while the honest one requires 707.8s. The attacker's mining efficiency is 0.001458 while the honest one is 0.001413. The attacker has a certain advantage.
	
	\item {\bf II}. When the attacker's hashrate is three times that of the honest one, the average time for the attacker to mine  a block is 221.9s, while the honest one takes 733.2s. The attacker's mining efficiency is 0.001502, and the honest miner is 0.00136. Also, the higher attack hashrate is, the more efficiency is.
\end{itemize}

\begin{figure*}[ht]
	\centering
	\subfigure[Equal Hashrate]{
		\includegraphics[width=8cm]{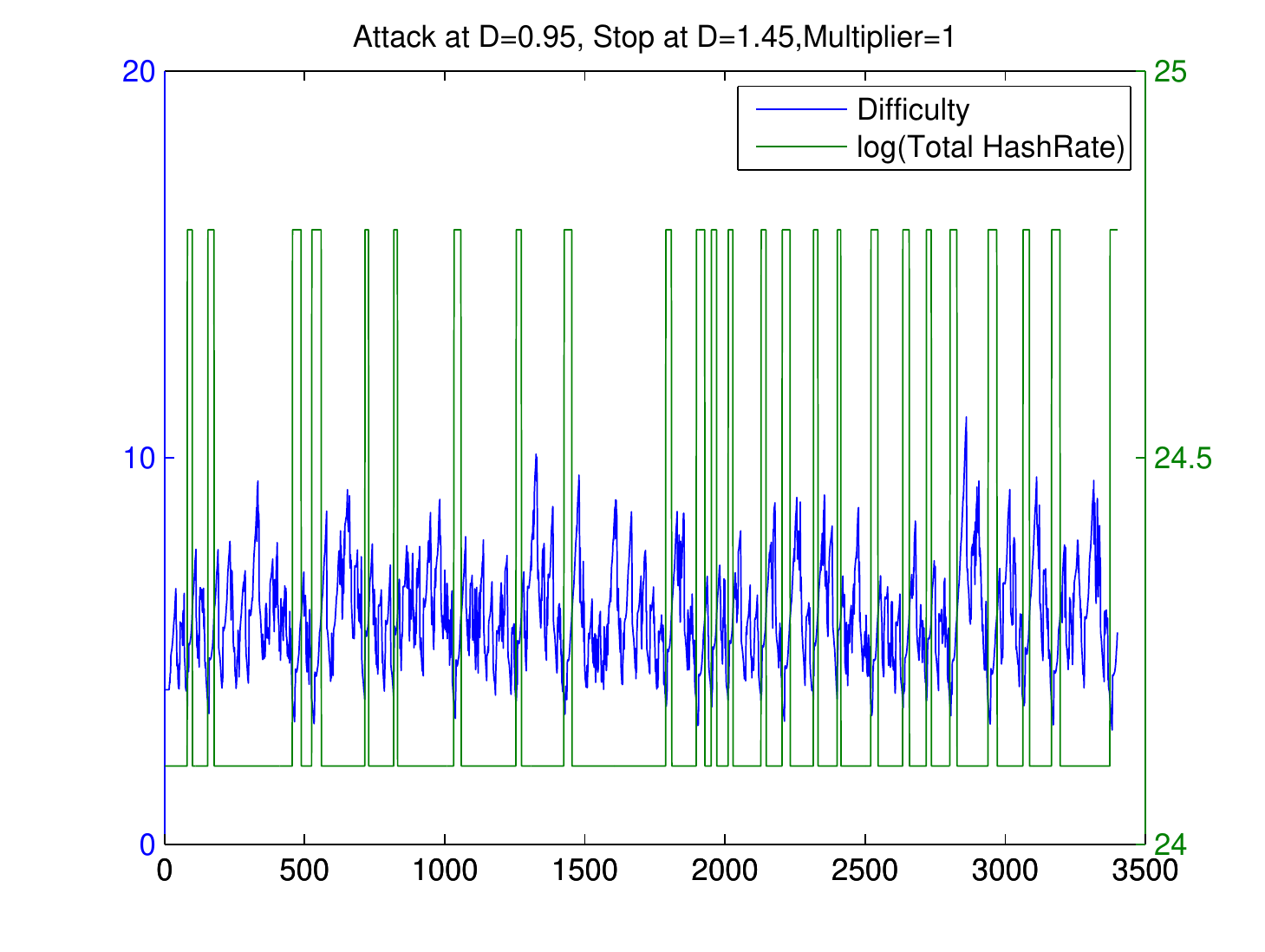}}
	\hspace{0in}
	\subfigure[Three Times Hashrate]{
		\includegraphics[width=8cm]{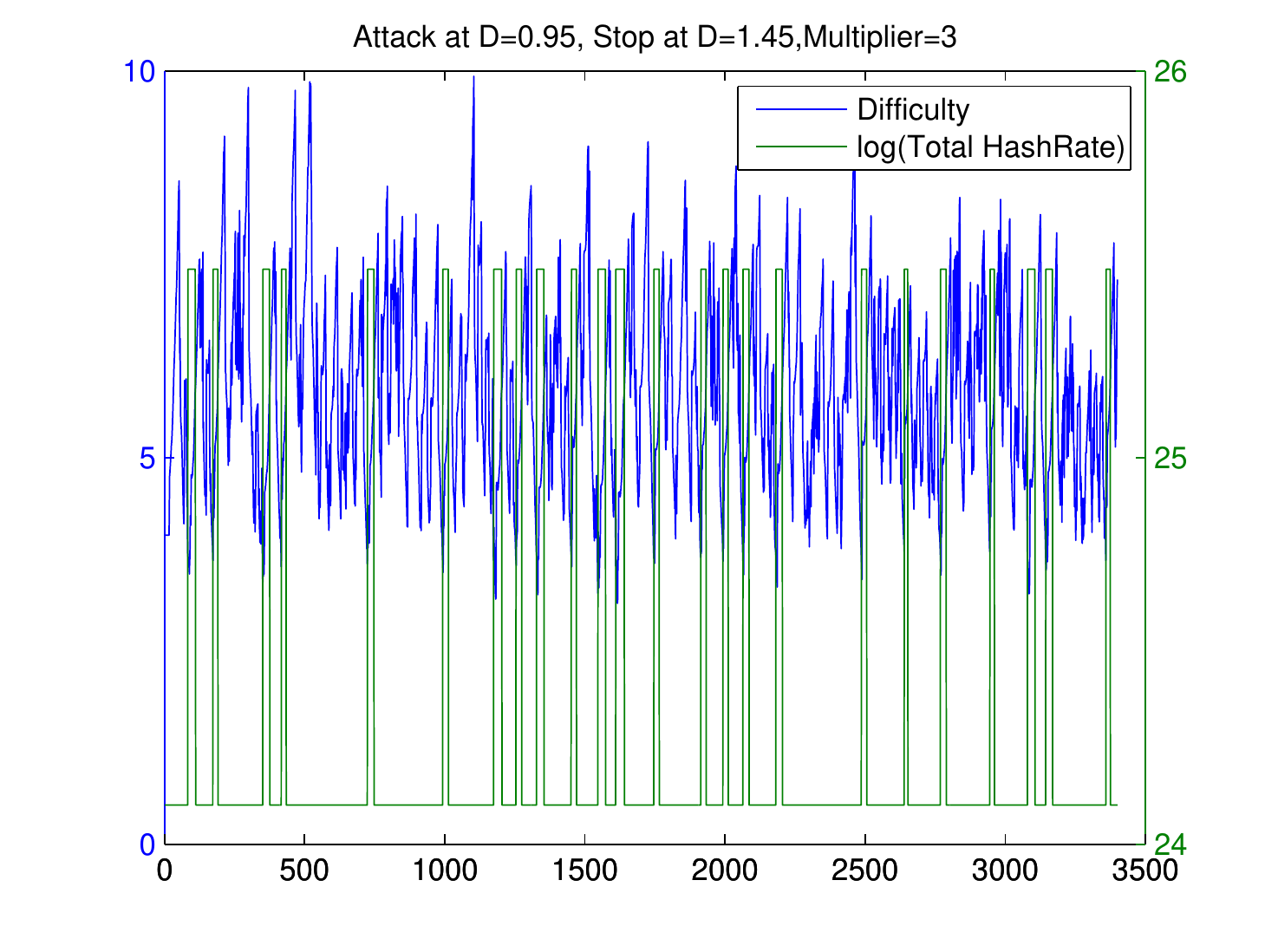}}
	\caption{\scriptsize{Simulation Results on Zcash}}
	\label{zcash}
\end{figure*}

\textbf{Results On ZCash}. The attack is still valid and the unfair situation gets worse as shown in Figure \ref{zcash}.

\begin{itemize}
	\item {\bf I}. When the attacker's hashrate is equal to the honest miner, the average time for the attacker to mine a block is 178.0s, while the honest one  requires 221.2s. The attacker's mining efficiency is 0.005617 while the honest one is 0.004521. The attacker has a large advantage.
	
	\item {\bf II}. When the attacker's hashrate is three times that of the honest one, the average time for the attacker to mine  a block is 54.9s, while the honest one takes  217.4s. The attacker's mining efficiency is 0.006070, and the honest miner is 0.004601. The higher attack hashrate is, the more efficiency is.
\end{itemize}

\begin{figure*}[ht]
	\centering
	\subfigure[Equal Hashrate]{
		\includegraphics[width=8cm]{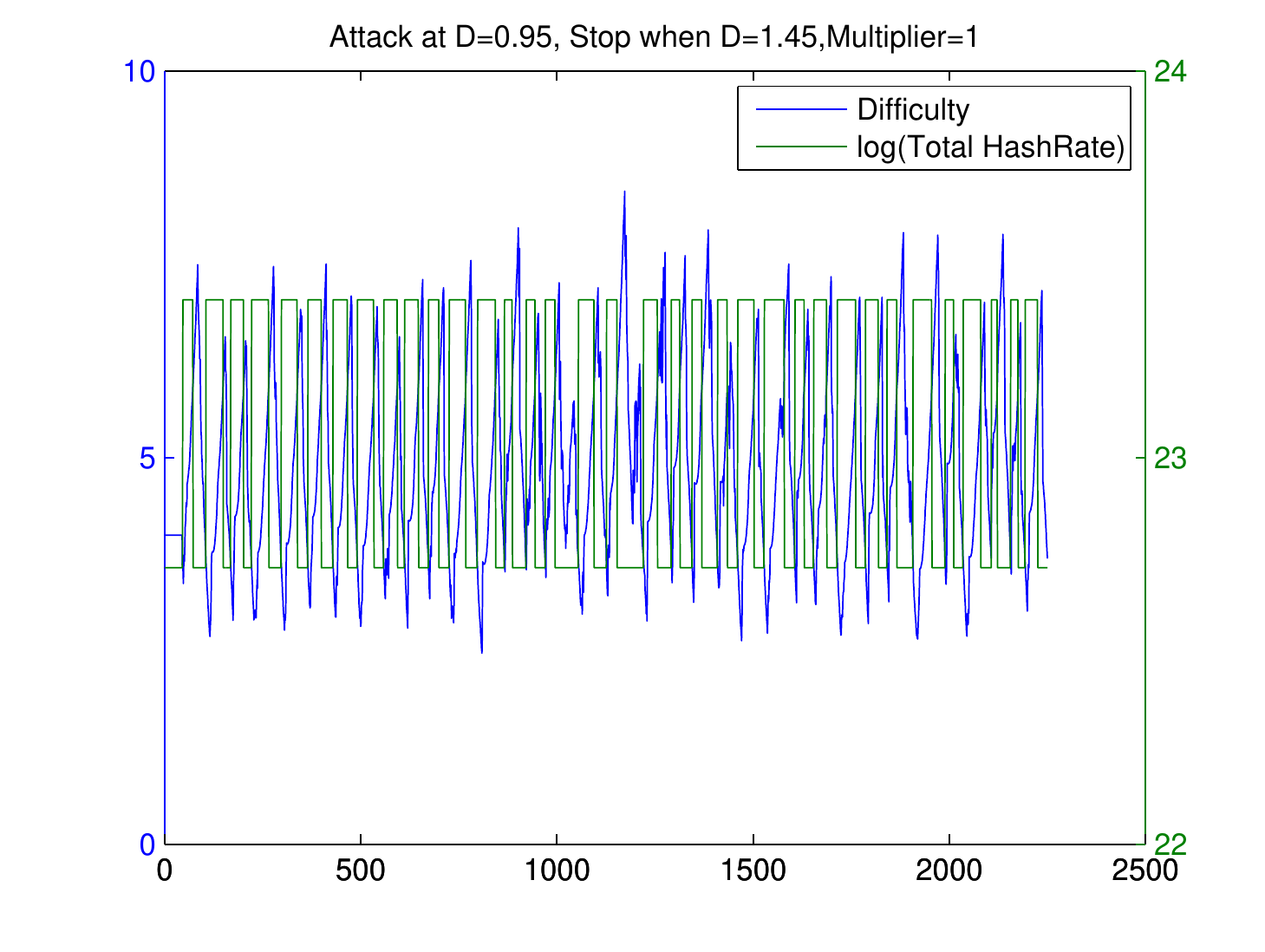}}
	\hspace{0in}
	\subfigure[Three Times Hashrate]{
		\includegraphics[width=8cm]{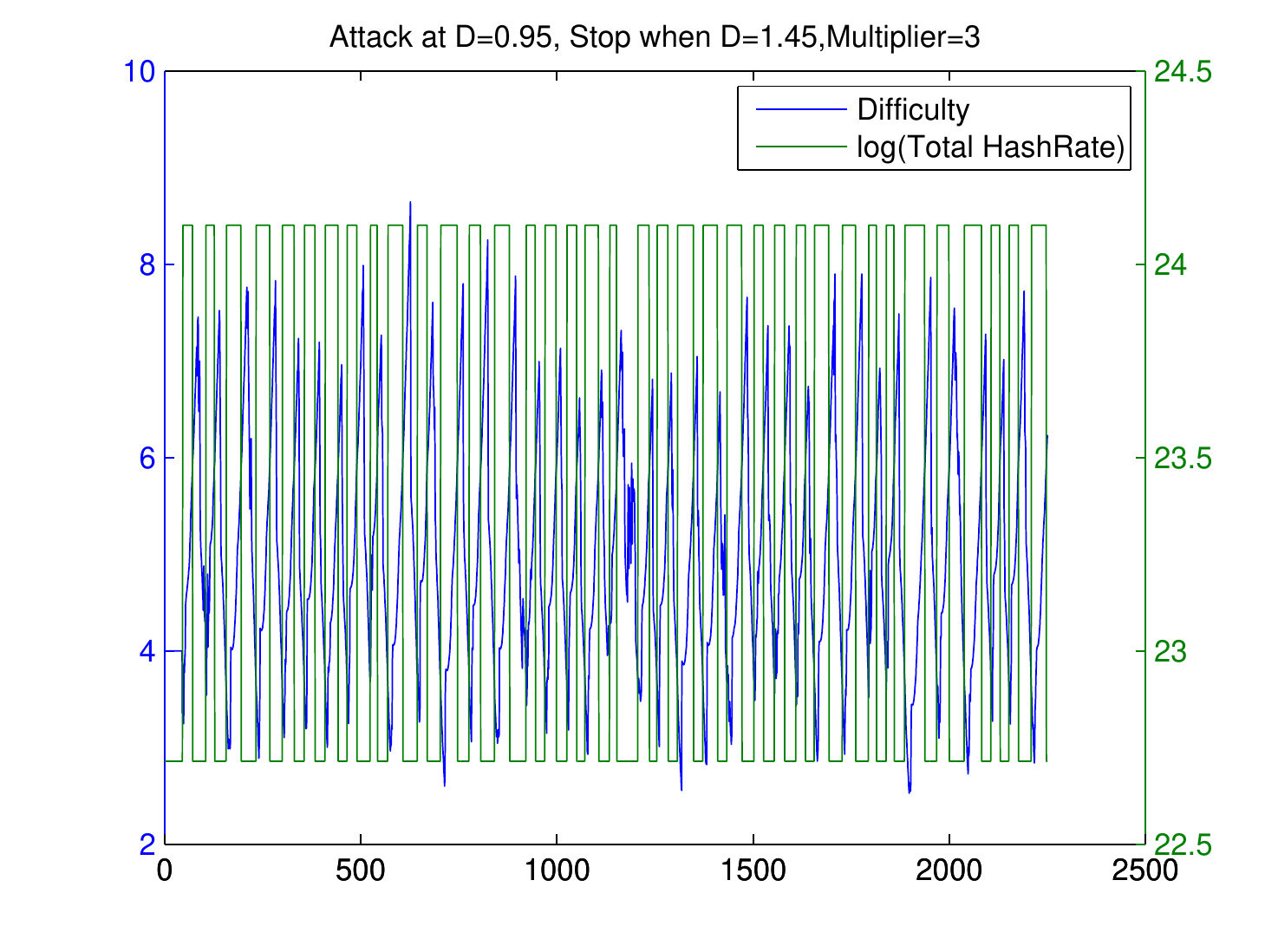}}
	\caption{\scriptsize{Results On Bitcoin Gold}}
	\label{attackbtg}
\end{figure*}

\textbf{Results On Bitcoin Gold}. A little better but the attack still works as shown in Figure \ref{attackbtg}.

\begin{itemize}
	\item {\bf I}. When the attacker's hashrate is equal to the honest miner, the average time for the attacker to mine a block is 667.0s, while the honest one  requires 771.5s. The attacker's mining efficiency is 0.001499 while the honest one is 0.001296. The attacker has a large advantage.
	
	\item {\bf II}. When the attacker's hashrate is three times that of the honest one, the average time for the attacker to mine  a block is 221.4s, while the honest one takes  794.0s. The attacker's mining efficiency is 0.001506, and the honest miner is 0.001259. The higher attack hashrate is, the more efficiency is.
\end{itemize}

\begin{figure}[htbp]
	\centering
	\includegraphics[width=9cm]{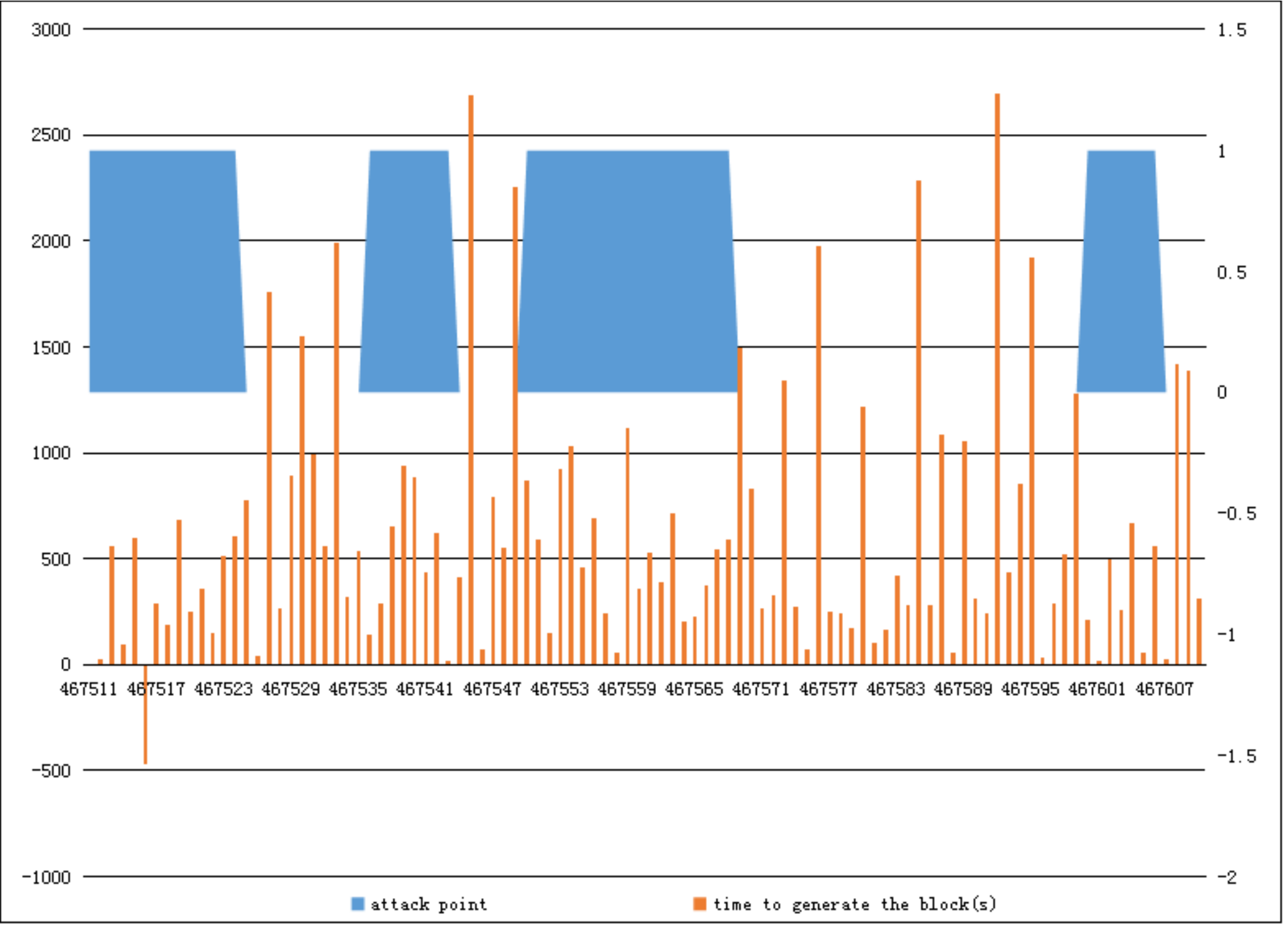}
	\caption{Attack Character from Bitcoin Gold's Public Block Data}
	\label{publicbtg}
\end{figure}

\textbf{Analysis on the public block data of Bitcoin Gold}. We selected about 150  blocks from the block data of the Bitcoin Gold \cite{btgdata}. As shown in Figure \ref{publicbtg}, the $x$ coordinate is the block height, and the $y$ coordinate is the block time. The orange histogram represents the generation time of different block heights, and the blue histogram represents the attackable area. We can see that there is a jumping mining attack on it from the character of the attack obviously.

\subsection{Summary}

From the experimental results, no matter which DAA above, the attacker can benefit from more or less by using the method of  jumping mining attack. What's worse is that when the attack hashrate increases, the attacker's efficiency get higher. This may lead to a result that honest miners of the blockchain network become less and less, and attackers are getting more and more.As a consequence, the foundation of the security of cryptocurrency may be shaken and occur the 51\% attack. In fact, BTG was attacked on May 18, 2018 \cite{btgnews}. The official website issued an announcement on May 24, 2018, admitting to being  attacked and explaining the situation and improvement plans. Of course, the actual motivation for the miner to mine needs to be calculated based on the market value of the cryptocurrency. Miners jumping to attack low-value cryptocurrencies may have lower income  than continuous mining on a high-value cryptocurrency. Further more, it is difficult for the attacker to gain the huge hashrate that  is multiples of honest miner in the whole network. That might be one of the reasons why we still haven't found similar attack on the BTC and BCH, the famous large cryptocurrencies using the existing DAA. However, the results tell us that this attack scheme is effective, and for new cryptocurrencies that are not protected by large computing hashrate, the aforementioned attacks are prone to happen.

\section{Anti-attack Scheme} \label{antiattack}

In this section, we further propose an effective difficulty adjustment algorithm for anti-jumping mining attack. Attackers are always looking for a way to find blocks with relatively low difficulty values in the block difficulty distribution curve for mining, while avoiding blocks with high difficulty as possible. Through the previous attack analysis and experiments in Section \ref{Attack}, we can see that an effective attack has at least the following characteristics:

\begin{itemize}
	\item {\bf I}. Attackers own a scale of computing hashrate.
	
	\item {\bf II}. When the attacker enters, the speed of adjusting the difficulty to increase  is not fast enough to stop the attacker to generate blocks burstly.
	
	\item {\bf II}. When the attacker jump out, the block difficulty started to adjust down but  the reaction was too intense. When the attacker leaves, the difficulty of the next block become high for the existing honest miners due to the decline in the computing hashrate of the entire network. If the adjustment continue to react too violently, it will be last a long time for the honest miners  to find  the next matching block. And then, the block difficulty will drops quickly again, thus the attacker can begin his next attack after several blocks generated.
\end{itemize}

\subsection{Our Improved DAA}

Based on the analysis above, we have improved the difficulty adjustment algorithm based on the weights of the block. Firstly, we continue to use the past $N$ blocks as feedback data for difficulty adjustment, and give the newer blocks a higher reference value by weight. After that, we separately monitor the block generation time  of the last 5 and 10 blocks to determine whether the computing hashrate of the entire network has suddenly increased, and adjust up the difficulty quickly. What's more, in order to prevent that the difficulty of generating blocks due to large delays decreasing rapidly, we limit the proportion of block difficulty adjusted down in the next block. In general, our solutions are to interfere with the attacker, making him unable to keep mining in the low difficulty level last a long time, and increasing the cost of his jumping mining. Thus, these reduce the profit of the jumping attacker. The key steps of our anti-attack DAA are described in Algorithm \ref{Antialgo}.


\renewcommand{\algorithmicrequire}{\textbf{Input:}}
\renewcommand{\algorithmicensure}{\textbf{Output:}}
\begin{algorithm}
	\caption{\textsf{GetNextDifficulty}}
	\label{Antialgo}	
	
	\begin{algorithmic}[1]
		\REQUIRE	\textit{DiffSeri}: the difficulty sequence of the last $N$ blocks;\\
		 \textit{STseri}: the sequence of the generation time of the last $N$ blocks;\\
		  $T$: average target time to generate a block.
		
		\ENSURE \textit{next\_Difficulty}: the difficulty of the next block.
		
		\FOR {$i=1$ to $N$, where $i$ stands for the block height}
		\STATE  \textit{solvetime} = \textit{STseri}($i$);
		\STATE  \textit{sum\_time} = \textit{sum\_time} + \textit{solvetime} $*$ $i$;
		\STATE  \textit{target} = \textit{getTarget}(\textit{DiffSeri}($i))$, here we have a \textit{target} = $\frac{2^{256-32}}{\textit{Difficulty}}$;
		\STATE  \textit{sum\_target} = \textit{sum\_target} + \textit{target};
		\IF {$(i \geq N-10+1)$}
		\STATE  \textit{sum\_last10\_time} = \textit{sum\_last10\_time} + \textit{solvetime}, record the time of the last ten blocks;
		\STATE  \textit{sum\_last10\_target} = \textit{sum\_last10\_target} + \textit{target};
		\ELSIF {$(i \geq N-5+1)$}
		\STATE  \textit{sum\_last5\_time} = \textit{sum\_last5\_time} + \textit{solvetime}, record the time of the last five blocks;
		\STATE  \textit{sum\_last5\_target} = \textit{sum\_last5\_target} + \textit{target};
		\ENDIF
		\ENDFOR
		\IF {$(sum\_time < N * N * T / 6)$}
		\STATE \textit{sum\_time} = $\frac{N * N * T}{6}$;
		\STATE keep \textit{sum\_time} reasonable in case strange \textit{solvetime} occurred.
		\ENDIF
		
		\STATE \textit{next\_target} =  $\frac{2 *sum\_time}{N * (N + 1)}$  * $\frac{sum\_target}{ N}$ * $\frac{adjust}{T}$, calculate the difficulty of the next block normally in the absence of attack;
		\STATE \textit{avg\_last5\_target} = $\frac{sum\_last5\_target}{5}$;
		\STATE \textit{avg\_last10\_target} = $\frac{sum\_last10\_target}{10}$;	
		\IF {\textit{sum\_last5\_time} $\leq 1.5*T$}
		\IF {$(next\_target >$ $\frac{avg\_last5\_target}{4}$)}
		\STATE \textit{next\_target} = \textit{avg\_last5\_target}/4;
		\ENDIF
		\ELSIF {\textit{sum\_last10\_time} $\leq 5 * T$}
		\IF {(\textit{next\_target} $>$ $\frac{avg\_last10\_target}{2}$)}
		\STATE  \textit{next\_target} = \textit{avg\_last10\_target}/2;
		\ENDIF
		\ELSIF {\textbf{ if}  $sum\_last10\_time \leq 10 * T$}
		\IF {(\textit{next\_target} $>$ $\frac{avg\_last10\_target*2}{3}$)}
		\STATE  \textit{next\_target} = $avg\_last10\_target*2/3$;
		\ENDIF
		\ENDIF
		\STATE \textit{last\_target} = \textit{getTarget}(\textit{DiffSeri}(\textit{end})); // the target of the previous block.
		\IF {$(next\_target >$ $\frac{last\_target * 13 }{10}$)}
		\STATE \textit{next\_target} = $\frac{last\_target * 13 }{10}$; in case difficulty drops too soon compared to the last block;
		\ENDIF	
		\IF {$(next\_target > pow\_limit)$}
		\STATE \textit{next\_target} = \textit{pow\_limit}; // \textit{pow\_limit} is the maximum value of \textit{PoW\_Target} set by the cryptocurrency.
		\STATE \textit{next\_Difficulty} = \textit{getTarget}(\textit{next\_target}).
		\ENDIF
		\STATE \textbf{return} \textit{next\_Difficulty}
	\end{algorithmic}
\end{algorithm}


\subsection{Attack test on the Improved DAA}

Similarly, we conducted extensive simulation experiments on the improved DAA. Through the experimental results, we can see that in the improved DAA, the attacker cannot take any advantage. When the attacker's hashrate is equal to the honest miner, the average time for the attacker to mine a block is 135.7s, while the honest miner's time requires 135.1s. The attacker's mining efficiency is 0.007369 while the honest one is 0.007403. The benefits of the honest miners are slightly higher than the attackers.

\begin{figure}[htb]
	\centering
	\includegraphics[width=8cm]{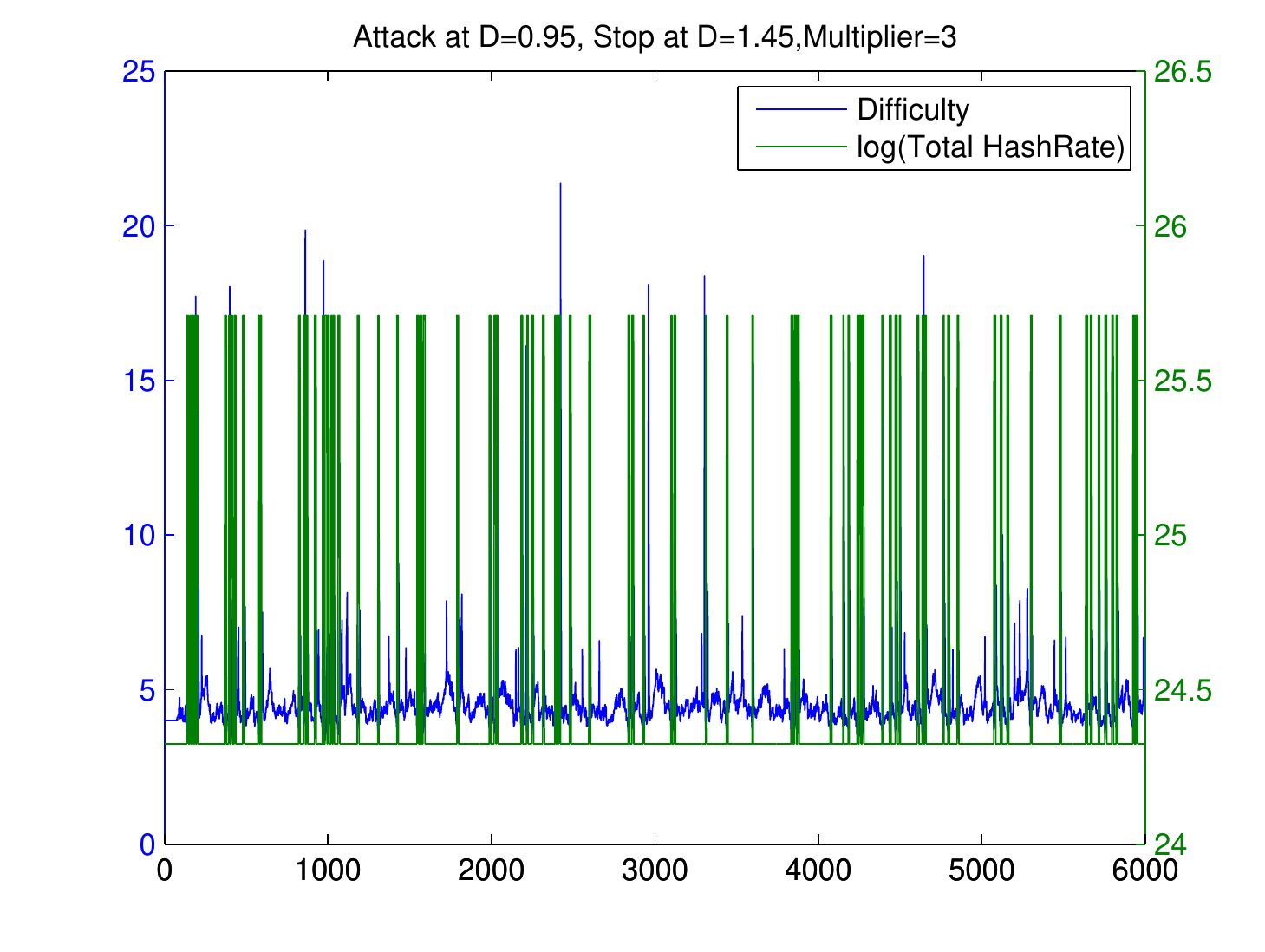}
	\caption{Attack test on the Improved DAA}
\end{figure}

\section{Conclusions} \label{conclusion}

Mining is one of the most critical part in the cryptocurrencies based on PoW. It determines the security of the blockchain on its consensus. Whether the profits of the miners match the hashrate they provide is crucial. If the honest miners devote their hashrate but cannot get a reasonable reward, then they will gradually decrease the hashrate in the network. And we know that hashrate is the basis for ensuring the security of the blockchain. The research of our paper starts from this key point. In the mining activities of the PoW blockchain, the difficulty adjustment algorithm  directly affects the income of the miners.

There are at least two limitations to analyze how the DAA of the public blockchain project works when the hashrate change: on the one hand, it takes a lot of time to generate enough block data for analysis even though we test it on testnet. On the other hand, it is hard to get enough hashrate for test. Therefore, a convenient and efficient research method is needed. We firstly propose a simulation model, which can effectively observe the relationship between network hashrate and block generation time. With this model, we analyze the DAA of several mainstream cryptocurrencies. By observing the character of these DAAs, we propose an attack scheme to make the attacker's income higher than the honest one. Furthermore, we conducted a large number of simulation experiments to verify the effectiveness of the attack scheme. In addition, we also analyzed BTG's historical block data to verify its existence of relevant attackable features. Finally, we propose an effective anti-attack scheme and also verify it through simulation experiments.

At present, our research work is still in its preliminary stage, and the following limitations still exist: (i) Our anti-attack method is only for jumping mining attack, and has not considered other mining attack schemes, such as selfish mining  \cite{eyal2014majority}; (ii) We have not fully considered the impact of the cost of an attacker's jumping and the market price of the cryptocurrencies on the behavior of miners. In future work, we would like to consider these factors and address these limitations.


\bibliography{main} 

\end{document}